\documentclass[revsymb,prb,twocolumn]{revtex4}%
\usepackage{graphicx}
\usepackage{bm}
\usepackage{slashed}
\usepackage{amsmath}
\usepackage{amsfonts}
\usepackage{wasysym}
\usepackage{amssymb}%
\usepackage{todonotes}
\providecommand{\U}[1]{\protect\rule{.1in}{.1in}}
\providecommand{\U}[1]{\protect\rule{.1in}{.1in}}
\newcommand{\be}{\begin{equation}}
\newcommand{\en}{\end{equation}}

\begin{document}
\title{Dark discrete breather modes in monoaxial chiral helimagnet with easy-plane anisotropy}
\author{I.G. Bostrem$^{1}$, E.G. Ekomasov$^{2,3}$, J. Kishine$^{4,5}$,  A.S. Ovchinnikov$^{1,6}$,  and V.E. Sinitsyn$^{1}$}
\affiliation{$^{1}$ Institute of Natural Science and Mathematics, Ural Federal University,
Ekaterinburg 620002, Russia}
\affiliation{$^{2}$ Bashkir State University, Institute of Physics and Technology, 
Ufa 450076, Russia} 
\affiliation{$^{3}$ Tyumen State University, Institute of Physics and Technology, 
Tyumen 625003, Russia} 
\affiliation{$^{4}$ Division of Natural and Environmental Sciences, The Open University of
Japan, Chiba 261-8586, Japan}
\affiliation{$^{5}$ Institute for Molecular Science, 38 Nishigo-Naka, Myodaiji, Okazaki,
444-8585, Japan}
\affiliation{$^{6}$ Institute of Metal Physics, Ural Division, Russian Academy of Sciences,
Ekaterinburg 620219, Russia}
\date{\today }

\begin{abstract}
Nonlinearity and discreteness are two pivotal factors for an emergence of discrete breather excitations in various media. We argue  that these requirements are met in  the forced ferromagnetic phase of the monoaxial chiral helimagnet CrNb${}_{3}$S${}_6$ due to specific domain structure of the compound. The stationary, time-periodic breather modes appear as  the discrete breather lattice solutions whose period mismatches with a system size. Thanks to easy-plane single-ion anisotropy intrinsic to CrNb${}_{3}$S${}_6$, these modes are of the dark type with  frequencies lying within  the linear spin-wave band, close to its bottom edge.   They represent cnoidal states of magnetization, similar to the well-known soliton lattice ground state,  with differing  but limited number of embedded $2\pi$-kinks. Linear stability of these dark breather modes  is verified by means of the Floquet analysis. Their energy controlled by two parameters,  namely the breather lattice period and amplitude,  falls off linearly  with a growth of the kink number.  These results may pave a new path to design spintronic resonators on the base of chiral helimagnets. 
\end{abstract}

\pacs{Valid PACS appear here}
\maketitle

\section{Introduction}

Nonlinear magnetic phenomena, where soliton-like waves are involved, have been extensively discussed for many decades, and a wealth of information has  been accumulated on the subject since that time \cite{Kosevich1990}.  Among various nonlinear coherent structures, such as kinks, vortices, or monopoles, breathers may appear as generic solutions of nonlinear dynamics  in a wide range of physical systems  \cite{Infeld1990}. Unlike  other soliton-like solutions, which preserve their shape as they propagate, the  breathers are characterized by spatial localization and periodic oscillations in time. This localization is realized in perfectly regular systems as opposed to disorder-induced Anderson localization.

When considering the discrete aspects of nonlinear lattice problems, the modified concept of discrete breathers (DB) is introduced \cite{Flach1998,Flach2008,Dmitriev2016}.  Discreteness is essential to preventing  resonances between the breather modes and the system characteristic frequencies.  Contrary to continuous solitons,  DBs do not require integrability for their
existence and stability. They are  not confined to a specific spatial dimension and have an amplitude dependent frequency. The  DB basic properties must be complemented by their division into the stationary and moving modes \cite{Bickham1997}.   These nonlinear excitations have been observed in a variety of physical systems,  where  discreteness arises either as an intrinsic property of the underlying lattice, such as   lattice dynamics of  halide-bridged transition metal complexes  \cite{Swanson1999} and  PbSe crystal \cite{Manley2019},  layered crystal insulator \cite{Russel2007},   or in artificially designed arrays of coupled Josephson junctions \cite{Binder2000,Trias2000}, coupled pendula \cite{Cuevas2009},  micromechanical cantilevers \cite{Sato2003},   optical waveguides \cite{Eisenberg1998}, in compressed  diatomic granular one-dimensional (1D) crystal \cite{Boechler2010},   and  in electrical circuit  with series-connected tunnel diodes \cite{Narahara2020}. 

As for DB excitations in spin lattices, they have been predicted for antiferromagnetic chains \cite{Lai1999}  with subsequent direct experimental  verification \cite{Schwarz1999,Sato2004}.  Currently, considerable numerical and theoretical progress has been achieved in understanding the nature of DB in Heisenberg spin chains with additional interactions, for instance, onsite anisotropy or next-nearest-neighbor exchange interactions\cite{Wallis1995,Rakhmanova1996,Zolotaryuk2001,Khalack2003,Lakshmann2014,Lakshmanan2018,Kamburova2019}.  Recently,  formation of  discrete breathers has been analyzed for weak ferromagnetic chains where the presence of the Dzyaloshinskii-Moriya (DM) interaction leads to a small canting between the interacting moments \cite{Djoufack2016,Kavitha2016}. The DB solutions have been examined in dynamics of the 1D array of magnetic particles (dots) with the easy-plane anisotropy and interparticle magnetic dipole interaction \cite{Pylypchuk2015}.

Most of realistic discrete systems may be reformulated for appropriate continuous media that often provide an  adequate description of nonlinear properties and, in some cases, analytical expressions in closed form may be derived \cite{Kosevich1990,Kosevich1974,Ivanov1979,Barjaktar1994,Papanicolaou1997}. However, it is quite possible that some nonlinear phenomena in continuous models can not appear in their discrete counterparts, and more comprehensive treatment is required to gain a deeper insight into the problem \cite{Gorbach2005,Quan2009}.    One  such issue is whether the discrete system  supports  breather lattice (BL) solutions, which represent a regularly ordered array of single breathers. These periodic solutions may be expressed in terms of Jacobi elliptic functions \cite{Byrd} and occur naturally for the continuous nonlinear equations, such as  the sine-Gordon  equation \cite{McLachlan1994,Kevrekidis2001,Fu2007a}, sinh-Gordon equation \cite{Fu2007c},   Korteweg--de Vries (KdV) equation \cite{Kevrekidis2003}, modified KdV equation \cite{Kevrekidis2004,Yan2008,Zhao2009},   nonlinear Schr\"odinger equation  \cite{Fu2007b}. Applications of the periodic breather modes to various continuous media have been discussed in Refs.\cite{Tankeyev2009,Tankeyev2010,Smagin2009,Yin2018}.  The BL solution has been predicted for continuous $\beta$-Fermi-Pasta-Ulam (FPU) chain\cite{Quan2008}, finding  periodic DB configurations in a triangular $\beta$-FPU lattice has been recently reported \cite{Babicheva2021}.  As for discrete spin lattices, spatially periodic breather type solutions have been  traced numerically in pioneering work \cite{Rakhmanova1998},  where they were named intrinsic localized spin modes of dark types by analogy with the optic solitons\cite{Weiner1988,Bao2018}.  However, to our knowledge, their analytical expressions are still lacking.

In this paper, we argue that the phase of forced ferromagnetism in the monoaxial chiral helimagnet CrNb${}_{3}$S${}_6$ assumes the existence of BL excitations. These modes are intrinsically stationary due to specific domain structure of CrNb${}_{3}$S${}_6$ samples  composed of 1 $\mu$m-wide grains with different, left- or right-hand,  crystallographic structural chirality \cite{Togawa2015}.  We demonstrate analytically and numerically that these spatially  periodic standing solutions with amplitude varying in time may be categorized into four types, depending  on the position their frequencies with respect  to the spectrum of linear spin waves.  This position is set by a value and sign of single-ion anisotropy, and, as a consequence, the only type of BL solutions, which is allowed by fairly small easy-plane anisotropy in CrNb${}_{3}$S${}_6$,    is the so-called dark breather modes emerging within spin-wave band, close to its bottom edge. They represent cnoidal states of magnetization, similar to the soliton lattice ground state,  with different number of embedded kinks.  The salient features of the ''bottom dark"  breather lattice are its period incommensurate with a system size and its energy falling off linearly with a growth of a number of  kinks. The linear stability of these modes is verified by means of Floquet analysis. 

It should be emphasized at the outset that previous theoretical investigations  of DB modes in ferromagnetic chains with on-site anisotropies \cite{Wallis1995,Rakhmanova1996,Rakhmanova1998} were inspired by  experimental studies of  spin dynamics in the linear chain compound  CsFeCl${}_3$ \cite{Schmid1994} for which intrinsic single-ion anisotropy exceeds intersite exchange coupling. A similar situation was dealt in Ref. \cite{Pylypchuk2015} where magnetic dots with a strong easy-plane anisotropy are coupled by weak magnetic dipole interaction. In contrast, our treatment is targeted at search of DB solutions for the case when exchange coupling between the nearest moments is much larger than on-site magnetic anisotropy.

The paper is organized as follows. In Sec. II, we describe a  model and give classification of periodic breather solutions. In Sec. III, we present results of numerical simulation of breather lattices based on discrete equations of spin motion. In Sec. IV, we present a detailed analysis of dark breather modes emerging near the bottom edge of the spin-wave spectrum. The conclusions and discussions are given in Sec. V.

\section{Model}

The model Hamiltonian of  the chiral monoaxial helimagnet is of the form
$$
\mathcal{H} = -2J \sum_n \textbf{S}_n \cdot \textbf{S}_{n+1}  + A \sum_n \left( S^z_n \right)^2   
$$
\begin{equation} \label{Hamilton}
+ D \sum_n \left[  \textbf{S}_n \times \textbf{S}_{n+1} \right]_z  - H  \sum_n S^z_n,
\end{equation}
where the first term is the exchange coupling along the chiral axis ($z$-axis) with  $J>0$. The second describes the single-ion anisotropy with the constant $A$, while the third does Dzyaloshinskii-Moryia interaction of the strength $D$. The last term denotes Zeeman coupling with an external magnetic field $H$ directed along the $z$ axis. The magnetic field is assumed to be strong enough, so that in the ground state all spins are ordered along the external field direction.  This forced ferromagnetic arrangement requires  $H >  2S \left( \sqrt{4J^2 + D^2} - J + A \right)$  \cite{Kishine2015}. 

The appropriate spin variables $s^{\pm}_n=\left( S^x_n \pm i S^y_n \right)/S$ and $s^z_n=S^z_n/S$  describe spin deviations from the ground state, where $S$ is the magnitude of spin. 

The equations of motion for these variables become
$$
\frac{i\hbar}{2JS} \frac{d}{dt} s^{+}_n =   
s^{+}_n \left(  s^z_{n+1} + s^z_{n-1} \right) - s^z_n \left( s^{+}_{n-1} +  s^{+}_{n+1} \right) 
$$
\begin{equation}  \label{Sraising}
  - 2B s^{+}_n s^z_n +  i \frac{D}{2J} s^z_n \left(  s^{+}_{n-1} -  s^{+}_{n+1} \right)  + \frac{H}{2JS} s^{+}_n, 
\end{equation}
where $s^z_n = \sqrt{1-\left|  s^{+}_n \right|^2}$ and $B=A/2J$.

Time-dependent solutions $s^{+}_n  = s_n(t) \exp \left( ikna - i\omega t \right)$ with  the wave number $k$  and the frequency $\omega$ are to be found,  where the amplitude $s_n$  is called {\it the transverse spin accumulation} and $a$ being the lattice constant. Substituting this form into Eq. (\ref{Sraising}) and collecting separately real and imaginary parts we obtain the system
$$
   \Omega s_n =  s_n \left(  \sqrt{1-s^2_{n+1}} + \sqrt{1-s^2_{n-1}} \right)  - 2B  s_n \sqrt{1-s^2_n}    
$$
\begin{equation}   \label{GenEq1}
-  \sqrt{1-s^2_n}  \left( s_{n-1} + s_{n+1} \right)  \sqrt{1+ \frac{D^2}{4J^2}}  \cos (ka + \delta), 
\end{equation}
\begin{equation} \label{GenEq2}
\frac{d s_n}{d \tau} = \sqrt{1-s^2_n}  \left( s_{n-1} - s_{n+1} \right) \sqrt{1+ \frac{D^2}{4J^2}}  \sin (ka + \delta),
\end{equation}
where $\tau = t/t_0$ is the dimensionless time with $t_0=\hbar/(2JS)$, the effective frequency $\Omega= \left( \hbar \omega - H\right)/2JS$, and the phase shift is determined from $\delta = \textrm{tan}^{-1} \left( D/2J \right)$. 

Breather spin modes are expected to exist in perfect discrete magnetic chains due to salient nonlinearity in the exchange and anisotropy interactions. These modes may lie inside the linear wave band  and splitting off from its upper or  lower edges, so-called DB solutions of the dark type \cite{Rakhmanova1998}.  By contrast, the bright type modes can appear at frequencies either just above the upper band edge or just  below the lower edge of the  linear spin wave spectrum.

 In the small spin deviation limit,  the linear  dispersion curve for spin wave is obtained from Eq.(\ref{GenEq1}) as  
\begin{equation} \label{SW}
\Omega (k) =  2    - 2B      -   2    \sqrt{1+ \frac{D^2}{4J^2}}  \cos ka. 
\end{equation}
The  top and bottom zone boundaries of the plane wave spectrum take the values  $\Omega (\pi)  =  2    - 2B     +   2    \sqrt{1+ \frac{D^2}{4J^2}}$, and   $\Omega (0) =  2    - 2B      -   2   \sqrt{1+ \frac{D^2}{4J^2}}$, respectively.

Concerning the DB wave number, it coincides with that of the simple or conical spirals in the static case when $d s_n/d \tau =0$, i.e. $ka = - \textrm{tan}^{-1} \left( D/2J \right)$. It immediately  follows  from Eq. (\ref{GenEq1}). 

To start with, we focus on the upper band edge and consider both bright and dark excitations. As  orientational order of  transversal spin components in these breather modes   is close to that of  the corresponding linear spin waves, i.e. to antiferromagnetic one,  it is convenient to introduce the envelope function $\psi (z) = (-1)^n s_n$, where $z=na$, which varies slowly in space. 

As will be seen in the next Section, the DB lattice modes may be compared with periodic solutions of the nonlinear Duffing equation\cite{Rakhmanova1996,Tankeyev2010}
\begin{equation} \label{Duffing}
  \frac{d^2 \psi}{d\tilde{z}^2} - \alpha \psi + \beta \psi^3 = 0,
\end{equation}
which can be deduced from Eq. (\ref{GenEq1}) if to keep only the leading cubic nonlinearity (for details, see Appendix A).   Here,  $\tilde{z}=z/a$ is the dimensionless coordinate expressed  in  lattice units,  and 
\begin{equation}  \label{alphaDef}
\alpha = \frac{\Omega - \Omega(\pi)}{\sqrt{1+\frac{D^2}{4J^2} } },   \quad 
\beta = - \frac{\Omega(\pi)}{2 \sqrt{1+\frac{D^2}{4J^2} }}. 
\end{equation}

For the case $\alpha >0$, $\beta >0$ the {\it bright}  BL solution centered at $\tilde{z}_0$ is obtained  as
\begin{equation} \label{BL1}
  \psi^{(+)}_{b} (\tilde{z}) =  \left[  \sqrt{ \frac{2c}{\beta} + \frac{\alpha^2}{\beta^2}} + \frac{\alpha}{\beta}  \right]^{\frac12}
\textrm{cn} \left[ \frac{4K}{L^{(+)}_b}  \left( \tilde{z} - \tilde{z}_0 \right), \kappa^2   \right],
\end{equation}
where $\textrm{cn}(\ldots)$ is Jacobi elliptic function with the modulus
 \begin{equation} \label{KappaTB}
 \kappa^2 = \frac12 \frac{\alpha + \sqrt{\alpha^2 + 2\beta c}}{\sqrt{\alpha^2+2\beta c}}.
\end{equation}
The constant  $c$ can take only positive values for these solutions that the superscript "+"  refers to.  Together with the requirement of smallness of  $ \psi^{(+)}_{b}$, it results in the condition 	
\begin{equation} \label{Cparam}
  0 < c < \frac{\beta}{2}-\alpha.
\end{equation} 
The constant $c$ specifies the amplitude and the bright  BL period  $L^{(+)}_b=4K\left( \alpha^2 + 2\beta c \right)^{-\frac14}$, where $K$ is the elliptic integral of the first kind.  For a finite size system $c$ may be derived from the requirement that dynamics of the edge and interior spins must be consistent (see Sec. III for details).

 In the opposite case $c<0$, the solution  is given by
 \begin{equation} \label{BB2}
  \psi^{(-)}_{b} (\tilde{z}) = \left[   \frac{\alpha}{\beta} - \sqrt{ \frac{2c}{\beta} + \frac{\alpha^2}{\beta^2}}  \right]^{\frac12} 
 \textrm{nd} \left[ \frac{2K}{L^{(-)}_b}  \left( \tilde{z} - \tilde{z}_0 \right) , \kappa^2   \right],
\end{equation}
where $\textrm{nd}(\ldots)$  is the Jacobi elliptic function $\textrm{dn}^{-1}(\ldots)$  with the modulus
 \begin{equation}
 \kappa^2 = \frac{2\sqrt{\frac{\alpha^2}{\beta^2}+\frac{2c}{\beta}}}{\frac{\alpha}{\beta}+\sqrt{\frac{\alpha^2}{\beta^2}+\frac{2c}{\beta}}},
\end{equation}
and the period $L^{(-)}_b=\sqrt{8}K \left[ \alpha + \sqrt{\alpha^2 + 2\beta c} \right]^{-\frac12}$.   In contrast to (\ref{Cparam}), the $c$-values are bounded from below $c > - \alpha^2/(2\beta^2)$. 

As expected, in the limit $ \kappa^2 =1$, when $c=0$, the bright  breather lattice (\ref{BL1}) reduces to the single breather solution 
\begin{equation}  \label{SingleT}
\psi_{b} (\tilde{z}) = \sqrt{\frac{2\alpha}{\beta}}  \frac{1}{\cosh \left[ \sqrt{\alpha} \left(  \tilde{z} - \tilde{z}_0 \right) \right]}
\end{equation}
with a peak shape of width $1/\sqrt{\alpha}$ centered around $\tilde{z}_0$, which can be loosely be thought as a "bound state" of a kink and antikink.  In contrast, the solution (\ref{BB2}) approaches zero when $\kappa^2$ goes to 1, but for all other $\kappa$ values the classification based on the number of kink-antikink bound pairs may be retained.

\begin{figure}[t] \label{ILM}
\includegraphics[scale=0.5]{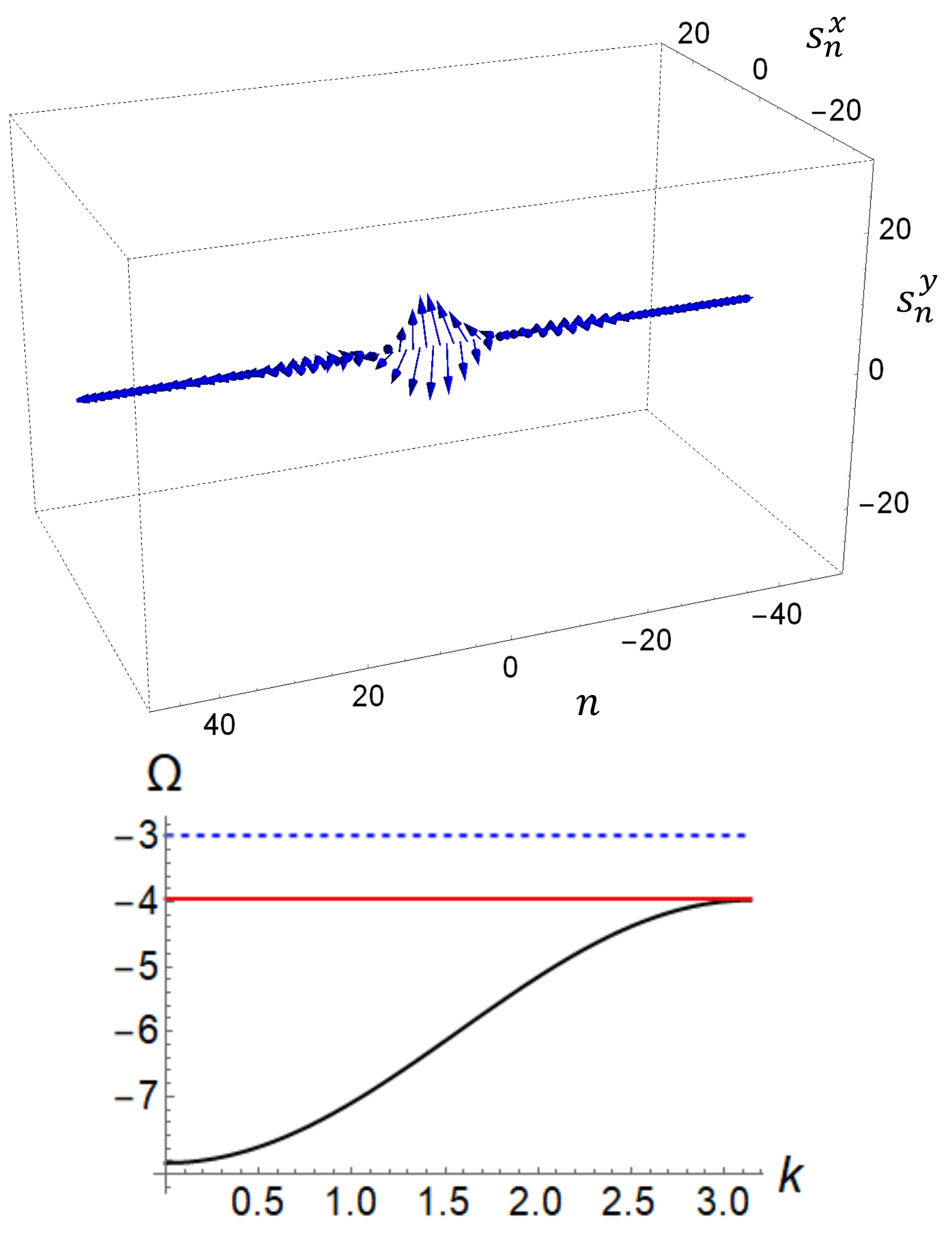}
\caption{\label{fig:epsart} The top bright  breather mode with one kink-antikink bound pair in the envelope function  built from numerical simulations  (upper panel).  Transversal spin components (relative to the chain axis) are shown with a factor 50 expansion.  Lower panel: the spin-wave spectrum at $B=4.0$ (black solid),  the breather frequency $\Omega=-3.95$ (red solid)  is limited from above  by $\Omega_u = \frac32 \left[ 1 -  B + \sqrt{1+ \left( D/2J \right)^2} \right]$  (blue dashed).}
\end{figure}

The numerical solution corresponding to the bright BL mode with the envelope function (\ref{BL1}) is presented in Fig.~1 (for details, see Sec. III).  It is known that the existence of such intrinsic localized modes (ILSM)  for a ferromagnetic chain with nearest-neighbor interactions requires that the strength of the single-ion anisotropy  exceeds a certain critical value so that the resulting ILSM frequencies can appear above the linear spin wave band \cite{Lai1999}.  In our case, the requirement $\beta>0$ imposes the restriction on the allowed anisotropy strength, $B > 1 + \sqrt{1+D^2/4J^2}$,  whereby it must exceed the Heisenberg exchange coupling. The condition is not suitable for the chiral helimagnet  CrNb${}_{3}$S${}_6$, where $B \sim 0.15$  \cite{Miyadai1983}  is close to the value for the antisymmetric exchange $D/2J \sim 0.16$. \cite{Shinozaki2016}   Notwithstanding, there are non-chiral magnetic materials, such as CsFeCl${}_3$  or some quasi-1D metal-organic compounds, where this somewhat exotic situation takes place. \cite{Anders1989}

\begin{figure}[t] \label{ILM}
\includegraphics[scale=0.5]{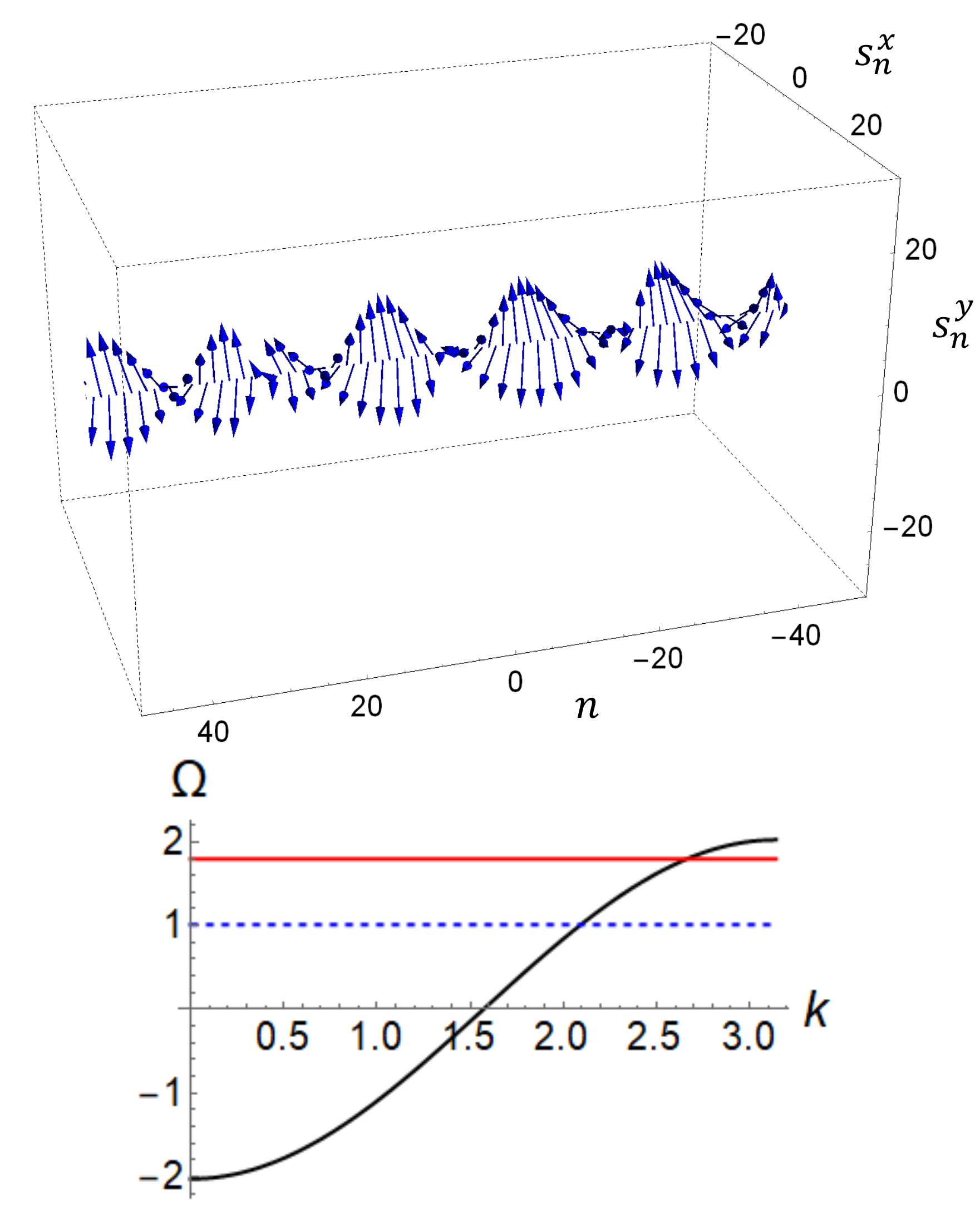}
\caption{\label{fig:epsart}    The top dark  breather mode  with 4 successive  kink/ antikinks in the envelope function  built from numerical simulations  (upper panel).  Transversal spin components (relative to the chain axis) are shown with a factor 20 expansion.  Lower panel:  Dispersion of the linear spin waves at the easy-plane anisotropy $B=1.0$ (solid black).   The breather frequency $\Omega=1.8$ is limited by the top edge of the spin-wave spectrum and $\Omega_d = 1 - B + \sqrt{1+ \left( D/2J \right)^2}$ (blue dashed).}
\end{figure}

The opposite situation, $\alpha <0$ and $\beta <0$, leads to the lattice breather solution of the {\it dark} type
\begin{equation} \label{BL2}
\psi^{(+)}_{d} (\tilde{z})  = \pm \sqrt{\frac{\alpha}{\beta} - \sqrt{\frac{\alpha^2}{\beta^2}+\frac{2c}{\beta}}}
\, \textrm{sn} \left[   
\frac{4K}{L_d} (\tilde{z} - \tilde{z}_0), \kappa^2
\right] 
\end{equation}
with the elliptic modulus
\begin{equation} \label{ElModDB}
\kappa^2 = \frac{\alpha + \sqrt{\alpha^2+2 \beta c}}{\alpha - \sqrt{\alpha^2+2 \beta c}}
\end{equation}
and the period $L_d=4K\left( -\alpha/2 +  \sqrt{\alpha^2/4+\beta c/2} \right)^{-\frac12}$.  Condition on the parameter $c$, which is positive,  repeats that of the bright modes (\ref{Cparam}).  Note that at $c<0$ only unbounded solutions appear what makes them   inappropriate as finite-amplitude excitations.

The corresponding single breather solution is retrieved when $\kappa^2 \to 1$, or $c \to -\alpha^2/(2\beta)$,
\begin{equation} \label{SingleB}
   \psi_{d} (\tilde{z}) = \pm \sqrt{\frac{\alpha}{\beta}} \tanh \left[  \sqrt{ - \frac{\alpha}{2}}  \left( \tilde{z} - \tilde{z}_0 \right)  \right].
\end{equation}
It describes a kink whose inverse width is $\sqrt{|\alpha|/2}$. Based on the form (\ref{SingleB}), breather lattices of the dark type may be categorized according to the number of consecutively embedded  S-shaped kinks/antikinks. 

An example of the dark type  DB excitations found numerically is shown in Fig. 2.  Its envelope function fits well with  the expression (\ref{BL2}) (see Sec. III for details).  Here we just note that the derivation of the dark breather solution is  based on the assumption $\alpha<0$ thereby meaning that the  effective frequency of the localized mode  cannot be upper  than the top of the spin-wave spectrum, $\Omega < \Omega(\pi)$ [see Eq. (\ref{alphaDef})]. Similarly to the case of bright modes, this fact results from the restriction on the single-ion anisotropy constant, $B < 1 + \sqrt{1+D^2/4J^2}$ originated formally from $\beta<0$. Although this estimate allows the {\it top dark} BL solutions (\ref{BL2}) for  fairly  small easy-plane anisotropy valid for CrNb${}_{3}$S${}_6$, the requirement $B>0$ turns out to be  incompatible  with smooth spin arrangement at boundaries (see Sec. III).

We now turn to the lower  edge of the spin wave band, where bright and dark DB modes may be also  found. Similar to the above classification, the bright localized excitations should  lie in the gap below the bottom of the spin-wave spectrum, whereas their dark counterparts may occur at frequencies above the bottom  spectrum edge. 

\begin{figure}[t] \label{ILM}
\includegraphics[scale=0.5]{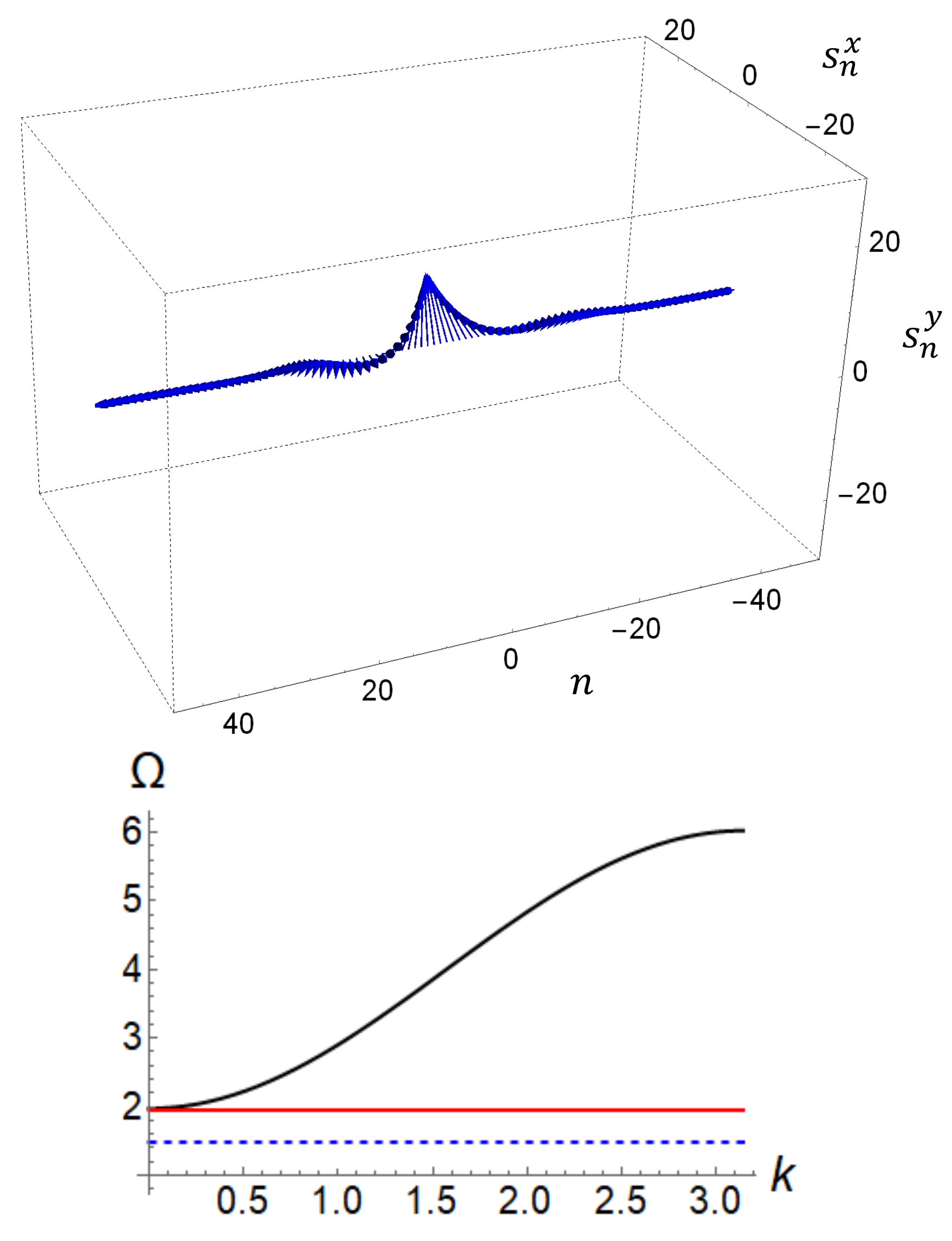}
\caption{\label{fig:epsart} The bottom bright breather mode  with one kink-antikink bound pair  in the envelope function  built from numerical simulations  (upper panel).  Transversal spin components (relative to the chain axis) are shown with a factor 50 expansion.  Lower panel:  Dispersion curve $\Omega(k)$ for the linear spin wave band (black solid) at the easy-axis anisotropy   $B=-1.0$ (solid black).   Frequency of the breather excitations $\Omega=1.95$ (red solid) with the lowest   boundary $\Omega_d = \frac32 \left[ 1 -  B -  \sqrt{1+ \left( D/2J \right)^2} \right]$  (blue dashed).}
\end{figure}

By invoking the continuum approximation the same Eq. (\ref{Duffing}) is recovered for the slowly varying  envelope $\psi_n = s_n$, but with different coefficients
\begin{equation} \label{alpha}
\alpha = \frac{\Omega(0)- \Omega }{\sqrt{1+\frac{D^2}{4J^2} } }, \quad 
\beta = \frac{\Omega(0)}{2\sqrt{1+\frac{D^2}{4J^2} }}.
\end{equation}  
Repeating the steps that led from Eq.(\ref{Duffing}) to DB modes, the  previous solutions (\ref{BL1},\ref{BB2}) and (\ref{BL2}) are recovered  as breather lattices of the bright and dark types, respectively. However, there is a significant discrepancy in conditions for $\Omega$ and $B$ owing to the different definition for  $\alpha$ and $\beta$.  As expected,  one gets $\Omega<\Omega(0)$ for the bright mode and  $\Omega > \Omega(0)$ for the dark mode what originates formally from  the requirements  $\alpha>0$ or $\alpha<0$, respectively.  From a magnetic viewpoint, the emergence of these excitations is related to the restriction imposed on the constant of the single-ion anisotropy $B$. The bright modes ($\beta>0$) appear at {\it easy-axis} anisotropy, $B<1- \sqrt{1+D^2/4J^2}$,  that rules out automatically excitations of this type for observation in  CrNb${}_{3}$S${}_6$.  Spatial distribution of magnetic moments of this type of DB mode, found numerically and corresponding to the model solution (\ref{BB2}), is illustrated in Fig. 3.  In addition to this, one may find a family of solutions (\ref{BL1}) for the same set of parameters.   In contrast,  the dark modes ($\beta<0$) comply with the opposite restriction  $B>1- \sqrt{1+D^2/4J^2}$, thereby admitting both fairly small easy-axis anisotropy ($B<0$) and that of the  easy-plane type ($B>0$).  Visualization of the spin arrangement for $B=0.15$ is given in Fig. 4.  It is precisely this type of DB excitations,  which can be approximated by the  {\it bottom dark} solution (\ref{BL2}),  is relevant to the chiral helimagnet CrNb${}_{3}$S${}_6$.

\section{Numerical simulations of  DB modes}

\begin{figure}[t] \label{ILM}
\includegraphics[scale=0.5]{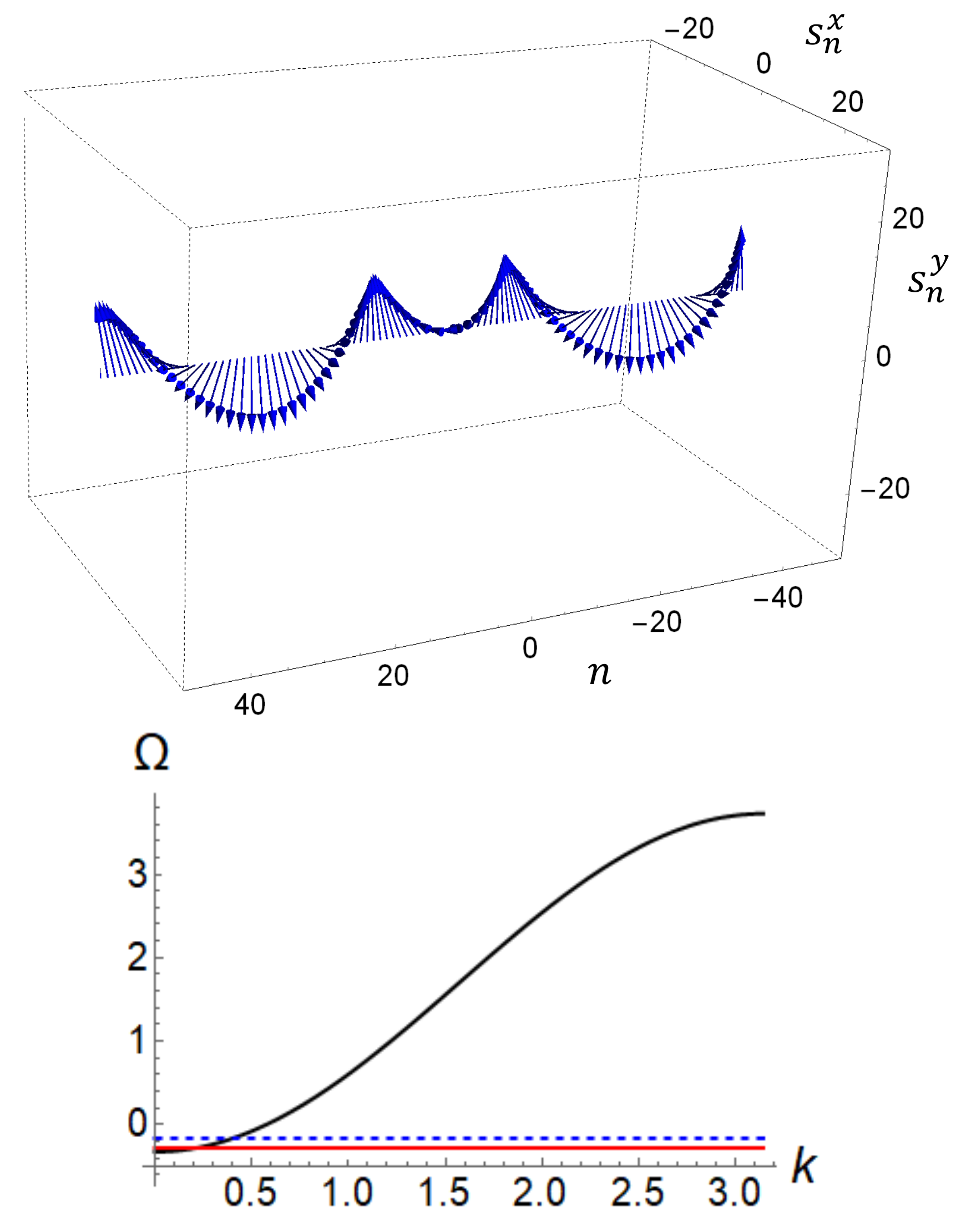}
\caption{\label{fig:epsart}  The bottom dark  breather mode with one kink  in the envelope function   built from numerical simulations  (upper panel).  Transversal spin components (relative to the chain axis) are shown with a factor 20 expansion.  Lower panel:  The spin-wave spectrum at the easy-plane anisotropy $B=0.15$ (solid black).  The breather frequency $\Omega=-0.28$ (red solid) is upper bounded by $\Omega_u =  1 - B - \sqrt{1+ \left( D/2J \right)^2}$ (blue dashed). }
\end{figure}

When spin deviations of the breather modes grow to be large enough, the basic equation (\ref{Duffing}) of the theoretical treatment becomes invalid. This warrants additional numerical simulations to verify analytical results. 

Our numerical procedure follows a scheme that has been suggested in Ref. \cite {Rakhmanova1996}. Giving an initial guess for $s_1$,  one can let $s_2$ be determined by the truncated version of Eq. (\ref{GenEq1}) 
$$
  \Omega s_1 =  - 2B  s_1 \sqrt{1-s^2_1}  +  s_1  \sqrt{1-s^2_{2}} 
  -  s_{2} \sqrt{1-s^2_1}     \sqrt{1+ q^2_0}
$$
with $q^2_0 =  \frac{D^2}{4J^2}$.

 After the initialization steps the general recursion takes the form
\begin{equation}
  s_{n+1} = \frac{-A_n \sqrt{\left( 1- s^2_n \right)\left( 1 + q^2_0 \right)}+\sqrt{1+\left( 1- s^2_n \right) q^2_0 - A^2_n}}{1+\left( 1- s^2_n \right) q^2_0}, 
\end{equation}
where
\begin{eqnarray*}
A_n  &=&  s_{n-1} \sqrt{1-s^2_n} \sqrt{1+ q^2_0}  \\
& & + s_n \left( \Omega \right.   \left.  - \sqrt{1-s^2_{n-1}  +2B \sqrt{1-s^2_n}}  \right).   
\end{eqnarray*}

The node index $n$ runs through the lattice, until it eventually reaches the right end. The truncated equation for the right edge spins
\begin{eqnarray*}
  \Omega s_L &=&    -  s_{L-1} \sqrt{1-s^2_L}     \sqrt{1+ q^2_0}  
\end{eqnarray*}
\begin{eqnarray} \label{Boundary}
&& + s_L  \sqrt{1-s^2_{L-1}} - 2B  s_L \sqrt{1-s^2_L}.
\end{eqnarray}
may be used as an inexpensive way to check convergence of the  iterative procedure.  Solutions are generated by scanning  $2 \cdot 10^{11}$ trial runs for $s_1$ from the range $[s_1, s_1+ h]$ with the step  $h=10^{-6}$. 

Examples of the four DB types, which are determined by their frequency positions  with respect to the spin wave band,  are illustrated in Figs. 1-4 (their temporal oscillations may be seen in Supplemental Materials). The chain length $L=101$  and the constant  DM strength $D/2J=0.16$ have been taken  in each case,    only values for $\Omega$, $B$ and $s_1$ varied.  In each of the plots, an additional line, bounding  possible frequency range of the breather excitations, is shown either  inside or outside the spin-wave band.  The boundary is established from the requirement that the amplitude of the appropriate single breather solution is limited to be less than 1. It can be easily seen from Eqs.(\ref{SingleT},\ref{SingleB}) that this is equivalent  to  the conditions $2\alpha < \beta$ or $\beta < \alpha$ for bright or dark modes, respectively. 

As such, only the  bright breather modes may be associated with self-localized spin-wave excitations (or, {\it intrinsic localized modes}), where a scale of the localization may be on a scale comparable to the lattice spacing. In the case of the bottom bright breathers,  the reason for this is that both the magnetic anisotropy and the magnetic field tend to align the moments all in the direction of the chain axis and only the DM interaction tries to prevent them.  For the top bright breathers, the single-ion anisotropy acts already in alliance with the antisymmetric  exchange coupling, but the interior antiferromagnetic order of moments restricts the spatial extent of these excitations.  

The latter plays a key role in a difference between  the dark breather solutions located at opposite ends of the spin-wave band. Although fairly small easy-plane anisotropy is permissible  for both of them there is an additional restriction on $B$ originated from the requirement that the frequency $\Omega$ of the interior spins must be the same as that of the end moments.

\begin{figure}[t] \label{TBB_EF}
\includegraphics[scale=0.9]{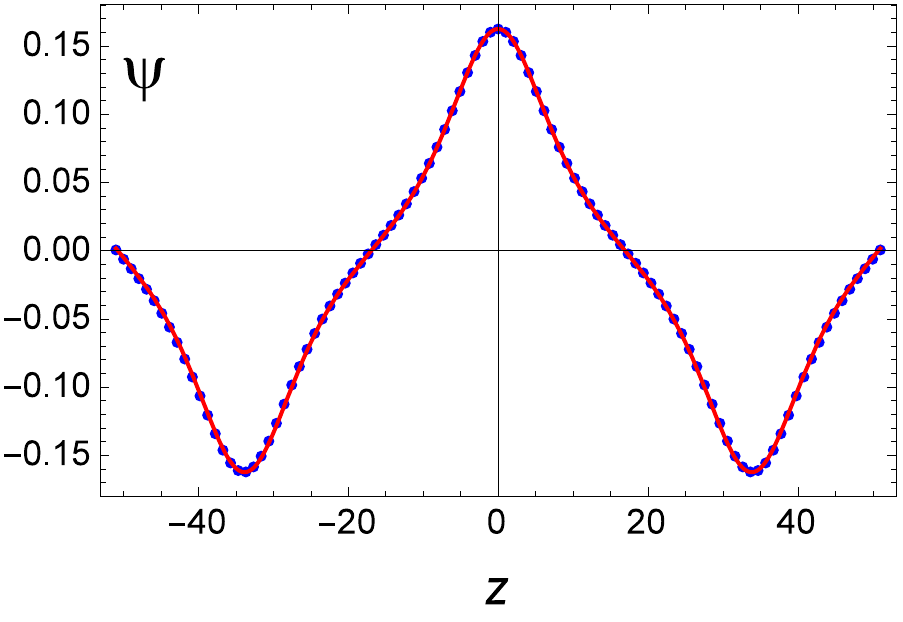}
\caption{\label{fig:epsart} The envelope function of the top bright breather lattice for $L=101$  as predicted by Eq.(\ref{BL1}). The solid red (dotted blue) line corresponds to analytical (numerical) calculation for $D/2J=0.16$ with $B=4.0$, $\Omega=-3.95$. The edge spin $s_{-L/2}=0.00048823$.}
\end{figure}

Indeed, the top dark breathers may arise at 
\begin{equation} \label{Omd}
\Omega > \Omega_d = 1- B + \sqrt{1+q^2_0}.
\end{equation}
On the other hand, Eq.(\ref{Boundary}) results in 
$$
  \Omega \approx  \sqrt{1-s^2_L}  \left( 1- 2B + \sqrt{1+q^2_0}  \right)
$$ 
provided the antiparallel alignment of the edge spins, $s_L \approx - s_{L-1}$.  Then, as an immediate consequence, 
\begin{equation}
 \sqrt{1-s^2_L} > \frac{1-B+ \sqrt{1+q^2_0}}{1-2B+\sqrt{1+q^2_0}}
\end{equation}
which leads to the obvious contradiction  $\sqrt{1-s^2_L} >1$ for  the easy-plane anisotropy $B>0$.

Frequency of the bottom dark breathers is bounded above by $\Omega_u = 1-B-\sqrt{1+q^2_0}$. Given ferromagnetic arrangement of moments at the edge, $s_L \approx s_{L-1}$, we have from  (\ref{Boundary})
$$
  \Omega \approx  \sqrt{1-s^2_L}  \left( 1- 2B - \sqrt{1+q^2_0}  \right).
$$

\begin{figure}[t] \label{BB_cn}
\includegraphics[scale=0.9]{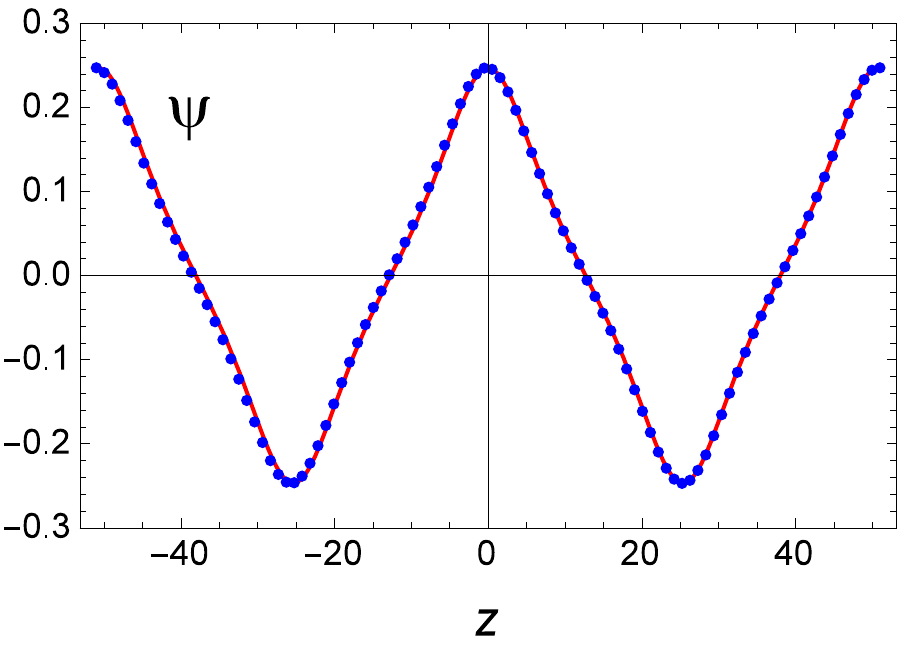}
\caption{\label{fig:epsart} The envelope function of the bottom bright breather lattice for $L=100$ as predicted by Eq.(\ref{BL1}). The solid red (dotted blue) line corresponds to analytical (numerical) calculation for $D/2J=0.16$ with $B=-1.0$, $\Omega=1.95$. The edge spin $s_{-L/2}=0.24708308$.
}
\end{figure}

Therefore,
$$
 \sqrt{1-s^2_L}   < \frac{B+ \sqrt{1+q^2_0}-1}{2B+\sqrt{1+q^2_0}-1},
$$
which amounts to the desired result  $ \sqrt{1-s^2_L}   < 1$  for any positive  $B$. 

In the foregoing analysis it was tacitly assumed that the spatial arrangement of the spin variables $s_n$ is smooth in the vicinity of the edges and  takes  zero values only inside the chain.     When, however,  $s_L$ is small enough,  the relationship 
$|s_L| \approx |s_{L-1}|$ is not longer applicable  and dark breather solutions become possible only under a suitable choice of $B$, ensuring matching of a spin configuration scale to the chain length.  That is the situation shown in Fig. 2. Moreover, solutions obtained thereby vary fairly rapidly with distance that makes the continuum approximation ineffective for them.

\begin{figure}[t] \label{BBB_EF}
\includegraphics[scale=0.9]{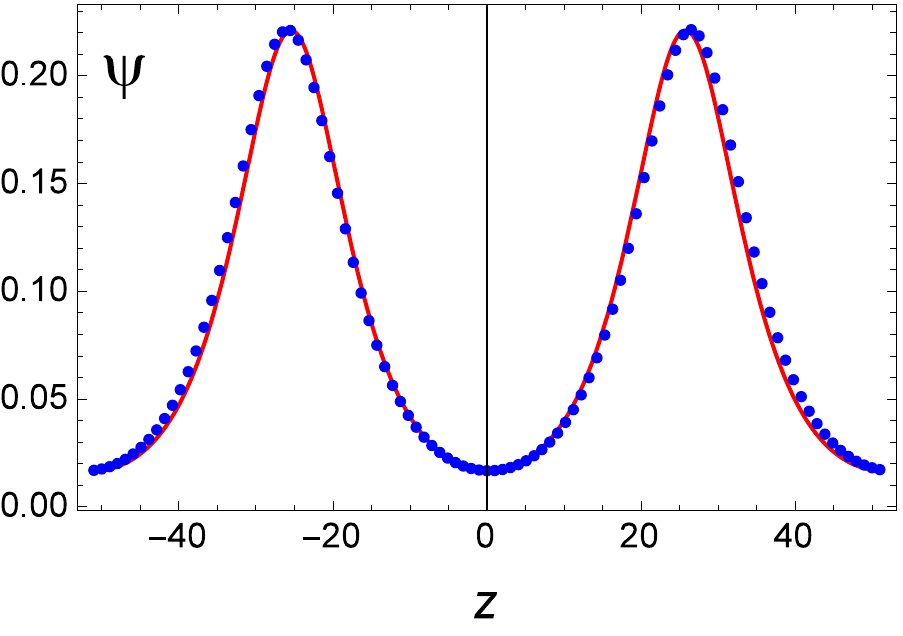}
\caption{\label{fig:epsart} The envelope function of the bottom  bright breather lattice for $L=101$ as predicted by Eq.~(\ref{BB2}). The solid red (dotted blue) line corresponds to analytical (numerical) calculation for $D/2J=0.16$ with $B=-1.0$, $\Omega=1.95$. The edge spin $s_{-L/2}=0.016876628$.}
\end{figure}

We shall now turn to fitting the analytical expressions of Sec. II to the envelope function $\psi(z)$ obtained numerically to confirm  the validity of the continuum approximation.  

Details of such a procedure  for the top bright breather lattices  have been discussed by us in Ref. \cite{Bostrem2021}.   The main purpose  is to find the $c$ parameter and then deduce the elliptic modulus $\kappa$ from Eq. (\ref{KappaTB}) with defined $\alpha$ and $\beta$.  Note that  in the case of the soliton lattice ground state,  the  constant $c$ is excluded from minimization of energy per unit length, since $c$  is related with the elliptic modulus $\kappa$. \cite{Dzyaloshinskii1964} 

By definition, the constant $c$ arises as the first integral of Eq. (\ref{Duffing}) 
\begin{equation} \label{FirstInt}
   c = \left( \frac{d \psi}{d \tilde{z}} \right)^2 - \alpha \psi^2 + \frac{\beta}{2} \psi^4.
\end{equation} 

On the other hand, the derivative of the envelope function at the
chain edge is given by
\begin{eqnarray}\label{Edge} 
 \left.  \left( \frac{d \psi}{d \tilde{z}} \right) \right|_{L/2} \approx  &&\frac{1}{\sqrt{1+q^2_0}} \left( 1+ \sqrt{1+q^2_0} - \Omega -2B \right) \psi_{L/2} \nonumber\\
&&+ \frac{1}{2\sqrt{1+q^2_0}} \left(2 B - 1 -  \sqrt{1+q^2_0} \right) \psi^3_{L/2},\nonumber\\
&&
\end{eqnarray}
which directly follows from Eq. (\ref{GenEq1}), if to account the open boundary conditions and neglect nonlinear terms including derivatives of the function $\psi$. 

It can be envisaged that the envelope function is small at edges of the chain due to intrinsic localization of the top bright  breathers. Then, Eqs. (\ref{FirstInt}) and (\ref{Edge}) result in the transcendental equation
$$
c \approx \left( \alpha^2 + \alpha \frac{2+\sqrt{1+q^2_0}}{\sqrt{1+q^2_0}} + 2 \frac{1+\sqrt{1+q^2_0}}{\sqrt{1+q^2_0}} \right) \psi^2_{L/2}.
$$
Comparison of the envelope functions computed both numerically and analytically from Eq. (\ref{BL1}) is presented in Fig. 5.  Note that the top bright breather modes may be both on-site centered \cite{Takeno1988} and intersite centered  \cite{Page1990}.    This type of solutions may be found below the spin-wave band as well (see Fig. 6), however, it requires the chain consisting of an even number of sites, since these modes are being intersite centered only.  A comparing procedure between the numerical data and the model solution  is similar to that of the dark  BL modes as explained below.

Fig. 7 shows the numerical data superimposed on the analytical curve for the bottom bright breather, as given by Eq. (\ref{BB2}). To  render the fit with known $\alpha$ and $\beta$, it is convenient to make use the property of the solution, $\psi_{\textrm{min}}/\psi_{\textrm{max}} = \textrm{dn} K = 1 - \kappa^2$, that yields $\kappa^2 \approx 0.994298$ in this case. Then, the argument coefficient in the Jacobi $\textrm{nd}$-function  is easily recovered as
$$
\frac{1}{\sqrt{2}}  \left(\alpha + \sqrt{ \alpha^2 + 2 \beta c} \right)^{\frac12} = \sqrt{\frac{\alpha}{2-\kappa^2}} \approx 0.155293.
$$

\begin{figure}[t] \label{TDB_EF}
\includegraphics[scale=0.9]{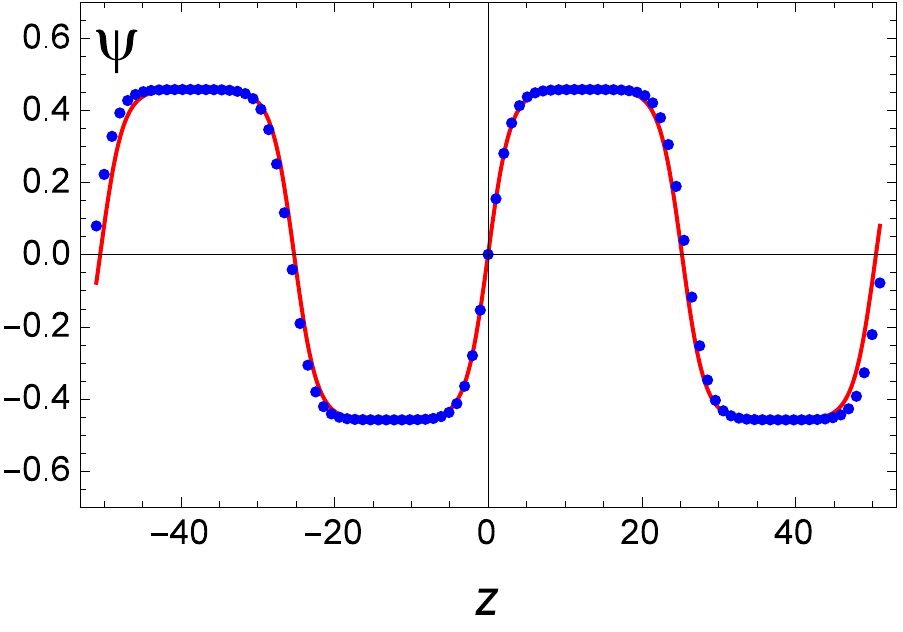}
\caption{\label{fig:epsart} The envelope function of the top  dark breather lattice for $L=101$ as predicted by Eq.~(\ref{BL2}). The solid red (dotted blue) line corresponds to analytical (numerical) calculation for $D/2J=0.16$ with $B=1.0$, $\Omega=1.8$. The edge spin $s_{-L/2}=-0.079802639$.}
\end{figure}

\begin{figure}[t] \label{BDB_EF}
\includegraphics[scale=0.9]{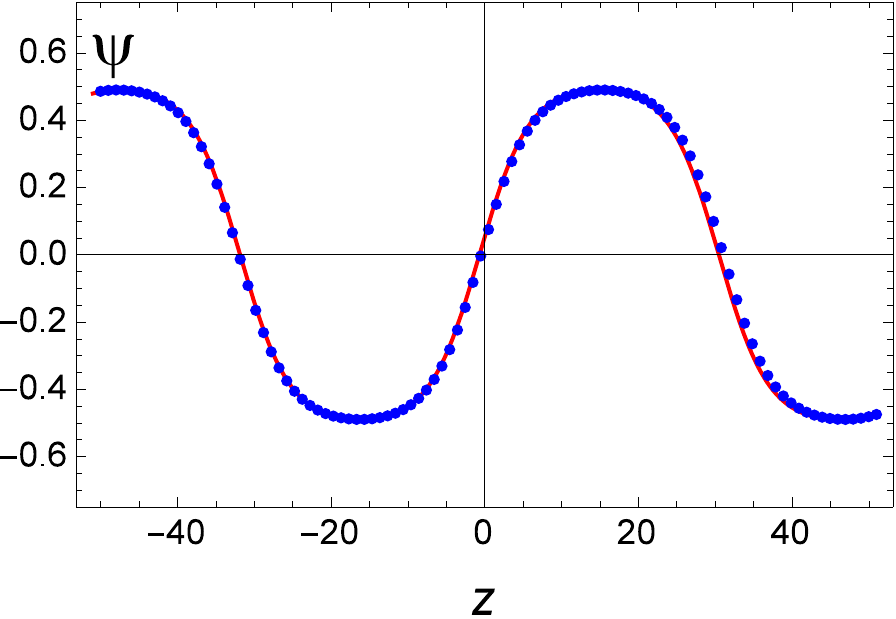}
\caption{\label{fig:epsart} The envelope function of the bottom  dark breather lattice for $L=101$  as predicted by Eq.~(\ref{BL2}). The solid red (dotted blue) line corresponds to analytical (numerical) calculation for $D/2J=0.16$ with $B=0.15$, $\Omega=-0.28$. The edge spin $s_{-L/2}=0.4858544687$.}
\end{figure}

It should be noted that although the top bright BL solutions (\ref{BB2}) are admitted  by the continuum theory, they are not reproduced numerically. This is probably related to antiparallel alignment of magnetic moments in these excitations.   

Fitting for the dark BL solutions  plotted in Figs. 8,9 follows the same template. At first, one finds the constant
$$
c = \psi^2_{\textrm{max}} \left(  |\alpha| - \frac{|\beta|}{2}\psi^2_{\textrm{max}}  \right),
$$
where $ \psi_{\textrm{max}}$ is maximum value of the envelope function adopted from numerical data. Next, the elliptic modulus (\ref{ElModDB}) and the  argument  in the Jacobi $\textrm{sn}$-function (\ref{BL2})  can be specified.

It is evident that predictions of the treatment based on the continuum approximation are in good agreement with their numerical counterparts.   From now on, solutions requiring either easy-axis or strong easy-plane anisotropy are beyond the scope of our study, and we narrow focus on the bottom dark breathers only as the most likely candidate to be  detected  in  CrNb${}_{3}$S${}_6$. 

\begin{figure}[t] \label{Floquet}
\includegraphics[scale=0.9]{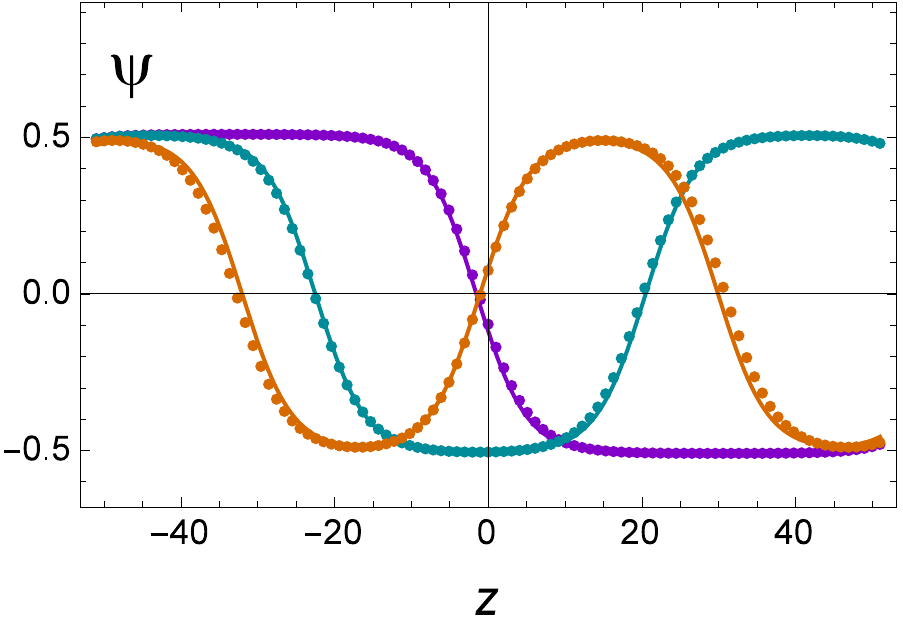}
\caption{\label{fig:epsart}  The envelope function of the bottom dark breather lattice  containing $N$ kinks:  $N=1$ (purple) with $s_{-L/2}=0.495373$, $N=2$ (teal) with $s_{-L/2}=0.494843$, $N=3$ (brown) with $s_{-L/2}=0.485854$. The other parameters as in Fig. 9.}
\end{figure}

\section{Bottom dark breathers}

To gain a deeper insight into the properties of the dark bottom breathers we start off with their classification.  Numerical simulations of these excitations for specific choices of the edge spin are depicted in Fig. 10. The main feature of the breather lattices is that their spatial period does not match the system size what resembles,  in some respects,  standing waves with soft pinning at boundaries \cite{Pincus1960,Kishine2019}.  Obviously, these solutions differ in a number $N$ of embedded kinks/antikinks, or, equivalently, by the number of nodes, that may be adopted as a criterion to categorize them. Finite spin deviations $s_{-L/2}$   needed to generate spin configurations of a given $N$  are slightly different from each other, but extremely high precision typical for top dark breather solutions  is not required.  

Next, we use the   Floquet theory  to examine linear stability of the identified breather modes \cite{Marin1998,Archilla2003,Khalack2003}. For the purpose, one has to study the evolution of a perturbation $\varepsilon_n (t)$  added to the DB periodic solution $s^{(0)}_n(t)$, i.e. $s_n(t) = s^{(0)}_n(t) + \varepsilon_n (t)$,  for the lattice of $L$ sites with $n=1,2,\ldots,L$.  Assuming that the size of perturbation is suitably small one may linearize the resulting equations for $\varepsilon_n (t)$ deduced directly from the equations of motion (\ref{Sraising}). Their explicit form looks a rather  cumbersome, but more importantly, the linearized system defines  the monodromy matrix $\hat{M}$, which maps $s_n(t)$ into $s_n(t+T)$,
\begin{equation} \label{Monodromy}
\left(
\begin{array}{c}
 \varepsilon^{'}_n(T) \\
 \varepsilon^{''}_n(T) 
\end{array}
\right)  =  \hat{\mathcal{M}}  
\left(
\begin{array}{c}
 \varepsilon^{'}_n(0) \\
 \varepsilon^{''}_n(0) 
\end{array}
\right).
\end{equation} 
 Here, the period  $T=\pi/\omega$  is only half the size of that of the periodic solutions  $s^{(0)}_n(t)$. In addition, the complex nature of $\varepsilon_n$ has to be taken into account to separate it into the real $ \varepsilon^{'}_n$ and imaginary $\varepsilon^{''}_n$ parts. 

\begin{figure}[t] \label{Floquet}
\includegraphics[scale=0.15]{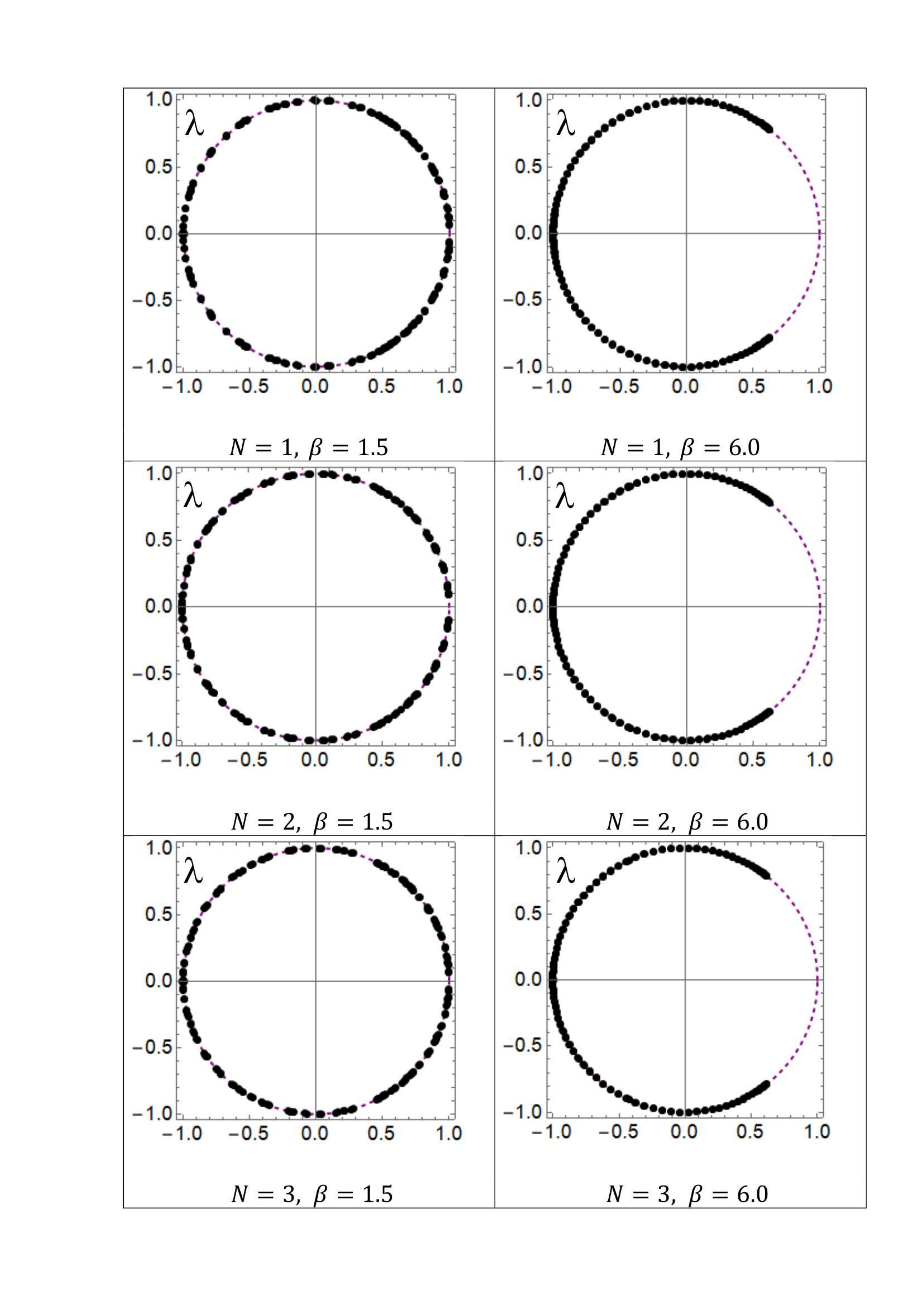}
\caption{\label{fig:epsart} Eigenvalues $\lambda$ of the monodromy matrix  for the bottom dark breather lattices of the size 101 with a different number of embedded kinks $N=1,2,3$ in  dependence on the magnetic field $\beta$.  The parameters are $B=0.15$, $\Omega=-0.28$, and $D/2J=0.16$. The critical magnetic field $\beta_{\textrm{cr}}=1.325$.}
\end{figure}

Any solution can  be determined by the column matrix of the initial conditions $\Lambda(0)=\left[\varepsilon^{'}_{1}(0),\varepsilon^{''}_{1}(0), \ldots, \varepsilon^{'}_{L}(0),\varepsilon^{''}_{L}(0) \right]^{T}$. The monodromy matrix can be easily constructed numerically by integrating  differential equations for $\varepsilon_n$  $2L$ times from $t=0$ to $T$ with the initial conditions  $\Lambda^{\nu}(0)$, $\nu=1,2,\ldots,2L$, with the elements $\Lambda^{\nu}_{\mu}(0) = \delta_{\nu\mu}$.  If the eigenvalues $\lambda$ of the monodromy matrix $\hat{\mathcal{M}}$  lie on the unit circle of the complex plane, then according to the Floquet theorem, the periodic orbit is stable,  otherwise it is unstable. 

Performing these calculations we obtain that the dark breather solutions are stable for  any $\beta > \beta_{\textrm{cr}}=2\sqrt{1+q^2_0}+2B-1$ (hereinafter,  $\beta=H/2JS$ is the magnetic field strength measured in the $2JS$ units). Close to the threshold point the Floquet eigenvalues are located almost uniform on the unit circle, while far from $\beta_{\textrm{cr}}$ they are redistributed being mostly concentrated on the left half of the unit circle, but stability nonetheless retains (Fig. 11).

\begin{figure}[t!] \label{Energy}
\includegraphics[scale=0.7]{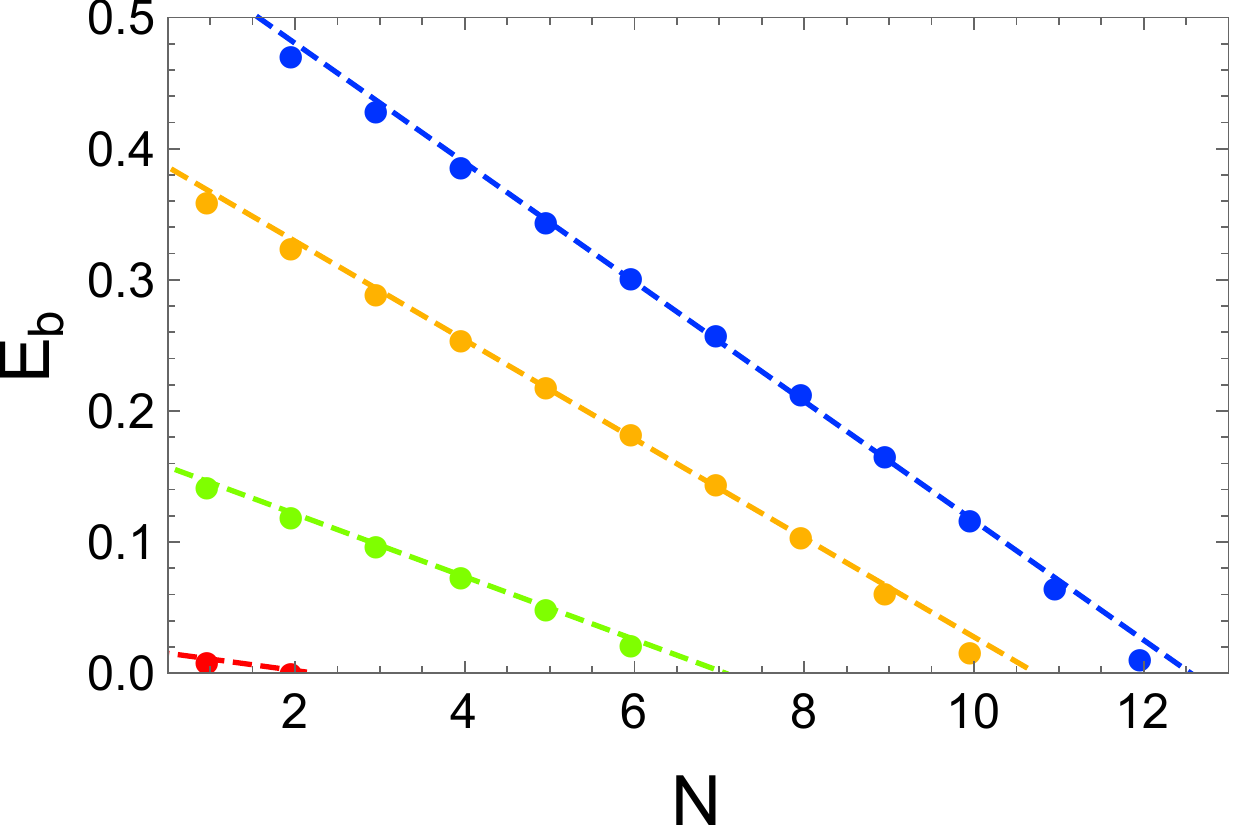}
\caption{\label{fig:epsart} Dependence of the BL energy per bond $E_b$ on the number of embedded kinks $N$ shown by dots for $\Omega=-0.18$ (blue), $\Omega=-0.22$ (brown), $\Omega = -0.28$ (green) and $\Omega=-0.32$ (red). Linear fitting to the numerical data is shown by the dashed lines. The parameters are $B=0.15$, $D/2J=0.16$ and $\beta=1.5$.  The frequencies are restricted to the range $\Omega(0) < \Omega < \Omega_u$ with $\Omega(0)= -0.325438$ and $\Omega_u = -0.162719$ (see the text) and corresponds to the physical  frequencies $\omega$ as  1.485 THz ($\Omega=-0.18$), 1.440 THz ($\Omega=-0.22$), 1.373 THz ($\Omega=-0.28$) and 1.328 THz ($\Omega=-0.32$). The chain length $L=101$.}
\end{figure}

An interesting feature of the bottom dark breather solutions is their energy $E$ dependence on the number  $N$ of embedded kinks. By rewriting the Hamiltonian (\ref{Hamilton}) in terms of the spin variables $s_n$ we get the form
$$
\frac{E}{2JS^2} = - \sqrt{1+ q^2_0} \sum_{n=1}^{L-1}  s_n s_{n+1} + B \sum_{n=1}^{L-1} \left( 1 - s^2_n \right) 
$$
\begin{equation}
-  \sum_{n=1}^{L-1} \sqrt{1-s^2_n} \sqrt{1-s^2_{n+1}} 
- \beta \sum_{n=1}^{L-1} \sqrt{1-s^2_n}.  
\end{equation} 
With the help  of the expression, the function $E_{\textrm{b}}$, the energy per a lattice bond measured from the ground state value $E_0=-1+B-\beta$, is plotted against $N$ in Fig. 12 for different values of the frequency $\Omega$ with $B$ and  $\beta$ fixed. 

To understand this result recall that the  orientation of the moments perpendicular to the chain axis is favorable from the viewpoint of easy-plane anisotropy, however, there is a significant increase in the Zeeman energy.  With an addition of a kink/antikink  the increase in the latter can be reduced in the vicinity of a kink center, however, this would give rise to an increase in the anisotropy energy.  The competition of the two contributions, whereof the Zeeman term dominates since the ratio $B/\beta$ is on the order of 0.1,  determines the details of dependence of $E_{\textrm{b}}$  on $N$, which is projected on a straight line with negative slope (Fig. 12). The exchange interaction plays no role owing to its full rotational invariance and the almost parallel alignment of adjacent magnetic moments. 

Being a continuous function of the frequency $\Omega$, the energy  $E_b$ decreases gradually to zero when $\Omega$ varies from the upper boundary $\Omega_u = 1- B - \sqrt{1+q^2_0}$ to the bottom edge of the spin-wave spectrum,  $\Omega(0) = 2-2B-2 \sqrt{1+q^2_0}$.  The requirement that $E_b$ should be nonnegative imposes a  restriction on a number of kinks in the breather lattice whereby the less breather frequency $\omega=2JS(\Omega+\beta)/\hbar$, the less maximal number of kinks. It should be mentioned at this point  that a similar trend has been experimentally observed in electrical lattices, where discrete breathers are produced via the well-known mechanism of modulation instability \cite{English2010,Palmero2011}.   Numerical  estimations for the physical frequency $\omega$ are given in Fig. 12  for $J=18$ K and $S=3/2$ relevant for CrNb${}_{3}$S${}_6$ \cite{Shinozaki2016}.  Furthermore, according to our calculations, the amplitude of the breather lattice modes becomes larger for the modes of higher frequency but with the same number of kinks. By definition, it corresponds to the regime of  hard nonlinearity. 

Finally, we point out that each of the discrete BL solutions possesses the topological charge $\mathcal{Q} =  {\Delta \varphi}/{2\pi} = {k L_0^{(d)}}/{2\pi}$,  where $\Delta \varphi$ is the total angle of spin moment rotation around the $z$-axis  per the BL period $L_d=4K\sqrt{(1+\kappa^2)/|\alpha|}$. This quantity includes the wave vector $k = - a^{-1} \tan^{-1} q_0$ resulting from the DM interaction. The same factor appears in the expression for the linear momentum density $\mathcal{P}_n=\hbar k S \left( 1 - \sqrt{1-s^2_n} \right)$ , which takes nonzero values whenever the transverse spin accumulation $s_n$ emerges. Thus, the important consequences of the DM interaction are topological protection ensured by the topological charge and a possibility of sliding motion, which requires no expense of energy,  provided by the nonzero linear momentum. 

\section{Results and Discussion.}

Noncollinear spin textures  in chiral helimagnets and their concomitant fundamental excitations  have aroused research interest as technologically relevant objects for spintronic applications \cite{Gobel2021,Inoue2021,Yang2020}.  Unfortunately, nonlinear breather excitations lie outside the mainstream of these investigations and the number of relevant works  is very limited \cite{Kiselev2012,Kiselev2013}.  In our study we argue that this kind excitations may arise in the form of magnetic discrete breathers in the monoaxial chiral helimagnet  CrNb${}_{3}$S${}_6$  owing to fairly small easy-plane anisotropy and the specific domain structure of the compound. The latter has strong impacts on standing spin waves in the thin films of CrNb${}_{3}$S${}_6$ \cite{Kishine2019}.  Namely the smallness of microdomains of definite crystallographic chirality necessitates discrete approach to studying the breather modes.  

We treat the phase of the forced ferromagnetic order, which is established above the threshold value of the external magnetic field directed along the chiral axis, and classify all types of possible solutions permitted by the model Hamiltonian of the monoaxial chiral helimagnet.  Using the data on magnetic anisotropy in  CrNb${}_{3}$S${}_6$, we estimate the relevance of these excitations   and conclude that only the dark breather  modes,  having the form of the periodic breather lattice,  are supported just above  the bottom edge of the spin wave spectrum.  This kind of nonlinear solutions, which is  proved to be stable,  supplements diversity of nontrivial spin structures in chiral helimagnets and may contribute to spin  response   in ESR experiments. 

The proposed classification scheme of DB solutions within  the continuum approach  offers a natural interpretation  of findings of  the early numerical studies on intrinsic localized spin modes in ferromagnetic Heisenberg chains \cite{Rakhmanova1996,Rakhmanova1998}.  It should be pointed out, however, that the inclusion of the Dzyaloshinskii-Moryia interaction is not simply a consideration for an additional form of  magnetic anisotropy, the antisymmetric exchange provides topological charge and linear momentum for emerging breather modes. 

In accordance with a generally accepted concept, DB modes owe their existence to two  pivotal components, discreteness and  nonlinearity.  In the case of  CrNb${}_{3}$S${}_6$, the discrete nature becomes evident by comparing the characteristic domain size  $L \sim$ 1  $\mu \textrm{m}$  with the BL period $L_0$.   Our calculations show (see, for instance, Fig. 10) that the period encompasses around 100 sites that yields $L_0 \sim$  0.1  $\mu \textrm{m}$,  i.e. only one tenth of $L$,  if the distance 6.847 $\AA$   between the nearest Cr ions along the chiral axis  \cite{Mandrus2013} is accounted for.    Nonlinearity of our problem is guaranteed by the single ion anisotropy $B$ and the constraint that the spin length is conserved in dynamical processes. 

The dark breather lattices may be classified by  the number $N$ of embedded kinks/antikinks, but modes with the same $N$ may appear at different frequencies going by the rule whereby the higher frequency, the bigger amplitude.    This fact imposes severe restriction upon  a maximum value of kinks/antikins that can be accommodated in a finite length segment. A striking feature of the dark breather lattice is its energy which decreases linearly with increasing $N$ in contrast to linear growth for bright modes \cite{Bostrem2021}.   This strange behavior reflects  a competition in the phase of forced ferromagnetism between the single-ion anisotropy, the antisymmetric exchange and the Zeeman coupling  whereof the last interaction  dominates and   what specifically makes the phase stable.  

These findings mean that the micron-sized domain of  CrNb${}_{3}$S${}_6$, in which the DB lattice is excited, is a system where areas of stored energy are linked to regions of transverse spin accumulation.      As  such, there arises a strictly 1D periodic array of resonators whose stored energy can take only discrete values and  is controlled by two degrees of freedom, namely,  the kink/antikink number $N$  (or equivalently, by the BL period) and   the BL frequency $\omega$ (or equivalently, by  the BL amplitude). According to our order-of-magnitude estimate, $\omega$ lies in the THz range.  Because of nonlinearity, an amount of stored energy in these "breather capacitors" will be far beyond the capacity of standing spin waves.  This functionality may be used to design spintronic resonators on the  base of chiral helimagnets.  Care should be taken to exclude resonances with linear spin waves, since frequencies of the dark breather modes lie inside the SW band. Whilst discreteness of the system ensures that these resonances may be avoided if the breather mode frequency is appropriately chosen,  this  issue requires more detailed examination in the framework of the special theory \cite{Kopidakis1999,Kopidakis2000}.  Another open issue relates to a way to excite in controllable manner  these breather modes  by external sources. The creation of special microfabricated microwave antennae, like those used in a noncentrosymmetrical ferromagnet LiFe${}_5$O${}_8$ to induce magnons with large momentum, appears to be a promising way \cite{Iguchi2015}. In this case, microwave wavelength emitted by the antenna must be matched with a period of the excited BL mode. These challenges posed by the practical application together with the basic problem of finding similar discrete breather modes in the another ordered phases of the monoaxial chiral helimagnet, i.e. conical and soliton lattice ones, should be addressed in future works.

\section{Conclusions.}

In summary, we investigate the possibility of  experimental observation of discrete breather excitations in the forced ferromagnetic phase of the monoaxial chiral helimagnet CrNb${}_{3}$S${}_6$. We found that the presence of the DM interaction does not prevent their emergence  but, conversely,  ensures their topological protection.  We demonstrate that the model Hamiltonian of the monoaxial chiral helimagnet   supports several kinds of possible spatially periodic discrete  breather solutions,  but  a specific choice depends on subtle interplay between interactions accounted for in the Hamiltonian. Using experimental data for strengths of the DM coupling and the easy-plane single-ion anisotropy in CrNb${}_{3}$S${}_6$  we predict for this magnetic material the appearance of the dark breather lattice modes with frequencies lying within the linear spin-wave spectrum near to its bottom edge. Results of numerical simulations of the DB excitations   on a finite-length   chain are in very good agreement with the continuum approximation, which relates the  DB lattice modes and  periodic solutions of the nonlinear Duffing equation. The Floquet analysis confirms stability of the discrete breather excitations  relevant for CrNb${}_{3}$S${}_6$. Their classification based on a number of embedded kinks/antikinks is suggested, and dependence of their energy   inversely proportional to this number is established. This unusual feature of the dark breather lattice modes positioned just above the spin wave band bottom  paves a new way  to design spintronic capacitors based on chiral helimagnets and operating on the principle energy pumping from the DB modes to the magnon sector. 

\begin{figure}[ht!] \label{PhaseTrajectories}
\includegraphics[scale=0.7]{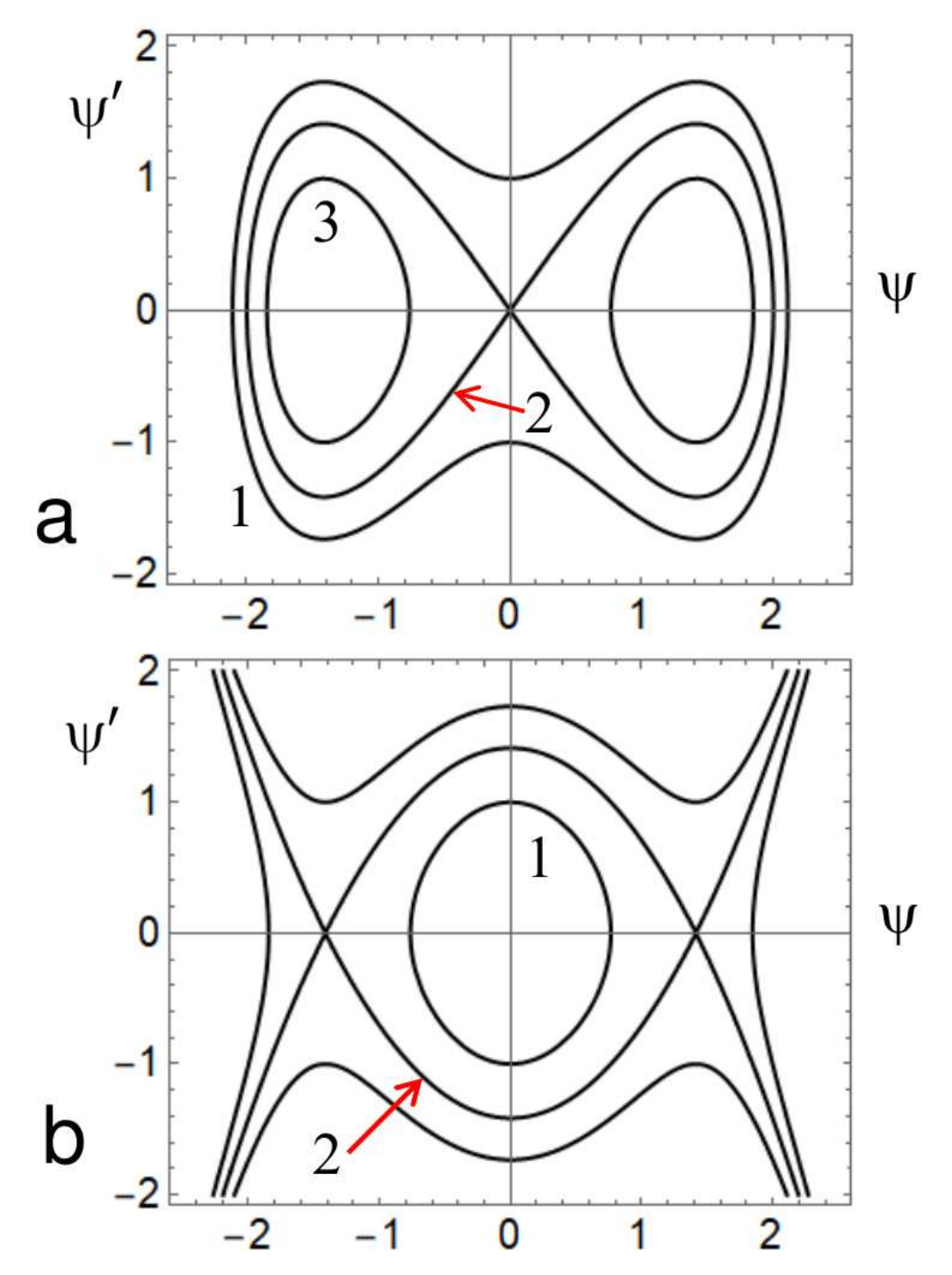}
\caption{\label{fig:epsart} Phase portraits of the Duffing equation (\ref{Duffing}). (a) The bright solutions at $\alpha=2$ and $\beta=1$; (b) the dark solutions at  $\alpha=-2$ and $\beta=-1$.}
\end{figure}

\begin{acknowledgments}
This work was supported by the  Act of the Government of the Russian Federation (contract  No. 02.A03.21.0006).  I.G.B.,  E.G.E. and V.E.S. acknowledge financial support by the Russian Foundation for Basic Research (RFBR), Grant No. 20-02-00213. A.S.O. thanks  the Russian Foundation for Basic Research (RFBR), Grant No. 20-52-50005, and the Ministry of Science and Higher Education of the Russian Federation, project No. FEUZ-2020-0054.  J.K. acknowledges financial support by   JSPS KAKENHI Grant No. 17H02923.
\end{acknowledgments}

\appendix

\section{Duffing equation}

In the longwave limit, when a characteristic spatial scale of DB solutions is much larger than the lattice unit $a$, one may proceed with the continuous coordinate $z=na$ instead of the discrete parameter $n$ in Eq. (\ref{GenEq1}).  

Approximating the spin variables by their continuous counterparts
\begin{equation}
s_n = (-1)^n \psi (z),
\end{equation}
\begin{equation}
s_{n\pm1} = (-1)^{n\pm 1} \left[  \psi(z) \pm  a \frac{d \psi}{dz}  + \frac{a^2}{2} \frac{d^2 \psi}{d z^2}  \right],
\end{equation}
\begin{equation}
 \sqrt{1-s^2_n}   \approx  1 - \frac12 \psi^2(z),
\end{equation}
\begin{equation}
\sqrt{1-s^2_{n\pm1}} \approx 1- \frac12  \psi^2(z) \mp a \psi(z) \frac{d \psi (z)}{dz},
\end{equation}
and  substituting them into Eq. (\ref{GenEq1}) we obtain eventually the  classical  Duffing equation (\ref{Duffing}) if nonlinear terms involving $d\psi/dz$ and $d^2 \psi/dz^2$ are neglected.

Behavior of  restricted solutions of the Duffing equation (\ref{Duffing}) is clearly illustrated by their trajectories on the phase plane $(\psi, d\psi/dz)$.  A type of the solutions depends on the signs of the coefficients $\alpha$ and $\beta$ (see, for example, Ref. \cite{Ostrovsky2002}).

 If $\alpha>0$, $\beta>0$ (Fig. 13a), the phase trajectories 1,2 and 3 conform with the breather solutions (\ref{BL1}), (\ref{SingleT}) and (\ref{BB2}), respectively. This is the case of the bright modes. 
 
 If $\alpha<0$, $\beta<0$ (Fig. 13b), the phase trajectories 1 and 2 correspond to   (\ref{BL2})  and (\ref{SingleB}), respectively. This is the case of the dark modes.

\end{document}